\documentclass[amsmath,aps,prd,showpacs,a4paper,10pt]{revtex4}

 \usepackage{epsf}
 \usepackage{graphicx}    

\begin{document}

 \newcommand{\be}[1]{\begin{equation}\label{#1}}
 \newcommand{\ee}{\end{equation}}
 \newcommand{\bea}{\begin{eqnarray}}
 \newcommand{\eea}{\end{eqnarray}}
 \def\disp{\displaystyle}

 \def\gsim{ \lower .75ex \hbox{$\sim$} \llap{\raise .27ex \hbox{$>$}} }
 \def\lsim{ \lower .75ex \hbox{$\sim$} \llap{\raise .27ex \hbox{$<$}} }

 \begin{titlepage}

 \begin{flushright}
 arXiv:0707.2129
 \end{flushright}

 \title{\Large \bf Age Problem in the Holographic Dark Energy Model}

 \author{Hao~Wei}
 \email[\,email address:\ ]{haowei@mail.tsinghua.edu.cn}
 \affiliation{Department of Physics and Tsinghua Center for
 Astrophysics,\\ Tsinghua University, Beijing 100084, China}

 \author{Shuang~Nan~Zhang}
 \affiliation{Department of Physics and Tsinghua Center for
 Astrophysics,\\ Tsinghua University, Beijing 100084, China\\
 Key Laboratory of Particle Astrophysics, Institute of High
 Energy Physics,\\
 Chinese Academy of Sciences, Beijing 100049, China\\
 Physics Department, University of Alabama in Huntsville,
 Huntsville, AL 35899, USA}

 \begin{abstract}\vspace{1cm}
 \centerline{\bf ABSTRACT}\vspace{2mm}
In this note, we test the original holographic dark energy
 model with some old high redshift objects. The main idea
 is very simple: the universe cannot be younger than its
 constituents. We find that the original holographic dark
 energy model can be ruled out, unless a lower Hubble
 constant is taken.
 \end{abstract}

 \pacs{95.36.+x, 98.80.Es, 98.80.-k}

 \maketitle

 \end{titlepage}

 \renewcommand{\baselinestretch}{1.6}



\section{Introduction}\label{sec1}
Dark energy~\cite{r1} has been one of the most active fields in
 modern cosmology since the discovery of accelerated expansion
 of our universe~\cite{r2,r3,r4,r5,r6,r7,r8,r9}. The simplest
 candidate of dark energy is a tiny positive cosmological constant.
 However, as is well known, it is plagued with the so-called
 ``cosmological constant problem'' and
 ``coincidence problem''~\cite{r1}. Many dynamical dark energy
 models have been proposed, such as quintessence~\cite{r10,r11},
 phantom~\cite{r12,r13,r14}, k-essence~\cite{r15},
 quintom~\cite{r16,r17,r18,r19}, hessence~\cite{r20}, and so on.

The so-called holographic dark energy is now an interesting
 candidate of dark energy, which has been studied extensively
 in the literature. It is proposed from the holographic
 principle~\cite{r21,r22} in the string theory. For a quantum
 gravity system, the local quantum field cannot contain too
 many degrees of freedom, otherwise the formation of black
 hole is inevitable and then the quantum field theory breaks
 down. In the thermodynamics of the black hole~\cite{r23,r24},
 there is a maximum entropy in a box of size $L$, namely the
 so-called Bekenstein entropy bound $S_{BH}$, which scales as
 the area of the box $\sim L^2$, rather than the volume
 $\sim L^3$. To avoid the breakdown of the local quantum field
 theory, Cohen {\it et al.}~\cite{r25} proposed a more
 restrictive bound, i.e. the energy bound. If $\rho_\Lambda$
 is the quantum zero-point energy density caused by a short
 distance cut-off, the total energy in a box of size $L$ cannot
 exceed the mass of a black hole of the same size~\cite{r25},
 namely $L^3\rho_\Lambda\,\lsim\, LM_{pl}^2$, where
 $M_{pl}\equiv (8\pi G)^{-1/2}$ is the reduced Planck mass.
 The largest IR cut-off $L$ is the one saturating the
 inequality. Thus,
 \be{eq1}
 \rho_\Lambda=3c^2M_{pl}^2L^{-2},
 \ee
 where the numerical constant $3c^2$ is introduced for
 convenience. If we choose $L$ as the size of the universe,
 for instance the Hubble scale $H^{-1}$, the resulting
 $\rho_\Lambda$ is comparable to the observational dark
 energy~\cite{r26,r27}. However, Hsu~\cite{r27} pointed out
 that in this case the resulting equation-of-state
 parameter~(EoS) is equal to zero, which cannot accelerate the
 expansion of our universe. The other possibility~\cite{r28}
 is to choose $L$ as the particle horizon
 \be{eq2}
 R_H=a\int_0^t\frac{d\tilde{t}}{a}
 =a\int_0^a\frac{d\tilde{a}}{H\tilde{a}^2},
 \ee
 where $H\equiv\dot{a}/a$ is the Hubble parameter; $a$ is the
 scale factor of the universe; a dot denotes the derivative
 with respect to cosmic time $t$. However, it is easy to find
 that in this case the EoS is always larger than $-1/3$ and
 also cannot accelerate the expansion of our universe~\cite{r29}.
 To get an accelerating universe, Li proposed in~\cite{r29} to
 choose $L$ as the future event horizon
 \be{eq3}
 R_h=a\int_t^\infty\frac{d\tilde{t}}{a}
 =a\int_a^\infty\frac{d\tilde{a}}{H\tilde{a}^2}.
 \ee
 In this case, the EoS of the holographic dark energy can be
 less than $-1/3$~\cite{r29}.

The theoretical aspects of the holographic dark energy have
 been studied extensively in the literature,
 see~\cite{r30,r31,r32,r33,r34,r35,r36,r37,r73,r74,r75,r76} for
 examples. Also, the holographic dark energy has been tested and
 constrained by various observations, such as SNe~Ia~\cite{r38},
 CMB~\cite{r39}, X-ray gas mass fraction of galaxy
 clusters~\cite{r40}, the differential ages of passively evolving
 galaxies~\cite{r41}, Sandage-Leob test~\cite{r42}, and so on.
 In particular, the holographic dark energy has been constrained
 recently by combining the latest SNe~Ia, galaxy clustering and
 CMB anisotropy~\cite{r43}. In the previous works, to our
 knowledge, the holographic dark energy is consistent with
 current observational data.

In this note, we test the original holographic dark energy with some
 old high redshift objects~(OHRO). The main idea is very simple:
 the universe cannot be younger than its constituents. In history,
 the age problem played an important role in the cosmology for
 many times. For example~\cite{r44}, before the discovery of
 accelerated expansion of our universe, many people believed that
 we are living in a matter-dominated spatially flat
 Friedmann-Robertson-Walker (FRW) universe. However, it is found
 that in this case, the present age of the universe,
 $t_0=(2/3)H_0^{-1}$~\cite{r45}, is smaller than the ages inferred
 from old globular clusters. The matter-dominated flat FRW universe
 is ruled out unless $h<0.48$, where $h$ is defined by the Hubble
 constant $H_0=100\,h\,{\rm km/s/Mpc}$. Since the age of the
 universe for the matter-dominated closed model is even smaller
 than the flat case~\cite{r45}, the age problem remains. Thus,
 for the matter-dominated FRW models without cosmological constant,
 only extremely open universe may be old enough to solve the age
 problem~\cite{r44}. The age problem becomes even more serious
 when we consider the age of the universe at high redshift (rather
 than at present day, $z=0$). By now, there are some OHRO are
 discovered, for instance, the 3.5~Gyr old galaxy LBDS 53W091 at
 redshift $z=1.55$~\cite{r46,r47}, the 4.0~Gyr old galaxy LBDS 53W069
 at redshift $z=1.43$~\cite{r48}. In addition, the old quasar
 APM 08279+5255 at $z=3.91$~\cite{r51,r52} is also used extensively.
 Its age is estimated to be 2.0--3.0~Gyr~\cite{r51,r52}.
 In~\cite{r53}, by using a different method, its age is reevaluated
 to be 2.1~Gyr. To assure the robustness of our analysis, we use
 the most conservative lower age estimate 2.0~Gyr for the old quasar
 APM 08279+5255 at $z=3.91$ throughout this work. In history, the
 former two old galaxies at $z=1.43$, $1.55$ have been shown to be
 incompatible with the age estimate for a flat FRW universe without
 cosmological constant. Combining with other independent observations,
 it is suggested that the more realistic model is the flat FRW
 universe with cosmological constant, i.e., the $\Lambda$CDM model.
 In the $\Lambda$CDM model, a period of cosmic acceleration at low
 redshift is allowed, and then the universe can have a larger age
 than the one of matter-dominated model. Therefore, the old galaxies
 at high redshift can be accommodated.

In fact, these three OHRO at $z=1.43$, $1.55$ and $3.91$ have been
 used to test many dark energy models, such as the $\Lambda$CDM
 model~\cite{r44,r53,r54}, the dark energy models with different EoS
 parameterizations~\cite{r49,r50}, the generalized Chaplygin
 gas~\cite{r55}, the $\Lambda(t)$CDM model~\cite{r56}, the
 model-independent EoS of dark energy~\cite{r57}, the scalar-tensor
 quintessence~\cite{r58}, the $f(R)=\sqrt{R^2-R_0^2}$ model~\cite{r59},
 the DGP braneworld model~\cite{r60,r61}, the power-law parameterized
 quintessence model~\cite{r62}, and so on. It is found that the two
 OHRO at $z=1.43$ and $1.55$ can be easily accommodated in most dark
 energy models, whereas the OHRO at $z=3.91$ can not, even in the
 $\Lambda$CDM model. These results give rise to the new age crisis
 in the dark energy models.

In this note, we consider the age problem in the original
 holographic dark energy model. In the next section, after a brief
 review of the holographic dark energy model, we compare the ages
 of these three OHRO with the ones estimated from the holographic
 dark energy model with the parameters constrained by the latest
 SNe~Ia, galaxy clustering and CMB anisotropy~\cite{r43}. We find
 that the two OHRO at $z=1.43$ and $1.55$ can be accommodated in
 the original holographic dark energy model, whereas the OHRO at
 $z=3.91$ can not. We then examine this age problem from another
 perspective. Following~\cite{r50}, we keep the model parameters
 $c$ and $\Omega_{m0}$ free, and find out the parameter space which
 can accommodate these three OHRO by means of plotting the contours
 of the dimensionless age parameter. The allowed parameter space
 can be ruled out at $1\,\sigma$ by the WMAP three-year~(WMAP3)
 bound $\Omega_{m0}=0.268\pm 0.018$~\cite{r6}. Even when we use
 the loosest and model-independent cluster estimate
 $\Omega_{m0}=0.3\pm 0.1$~\cite{r63}, the allowed parameter
 space becomes very narrow and can also be ruled out by the
 combined constraints from the latest SNe~Ia, galaxy clustering
 and CMB anisotropy~\cite{r43}. To alleviate the age problem in
 the holographic dark energy model, as shown in Sec.~\ref{sec3},
 a lower Hubble constant is needed, such as the one recently
 advocated by Sandage and collaborators. In the last section,
 some concluding remarks are given.


\section{The holographic dark energy model versus OHRO}\label{sec2}
The age of our universe at redshift $z$ is given
 by~\cite{r45,r50,r44,r49,r53,r54,r55}
 \be{eq4}
 t(z)=\int_z^\infty\frac{d\tilde{z}}{(1+\tilde{z})H(\tilde{z})}.
 \ee
 It is convenient to introduce the so-called dimensionless
 age parameter~\cite{r50}
 \be{eq5}
 T_z(z)\equiv H_0 t(z)=\int_z^\infty
 \frac{d\tilde{z}}{(1+\tilde{z})E(\tilde{z})}\,,
 \ee
 where $E(z)\equiv H(z)/H_0$. At any redshift, the age of
 our universe should be larger than, at least equal to, the
 age of the OHRO, namely
 \be{eq6}
 T_z(z)\geq T_{obj}\equiv H_0 t_{obj},
 ~~~~~~~{\rm or~equivalently,}~~~~~~~
 \tau (z)\equiv T_z(z)/T_{obj}\geq 1,
 \ee
 where $t_{obj}$ is the age of the OHRO. We consider a flat FRW
 universe which contains the holographic dark energy and
 pressureless matter. The Friedmann equation is given by
 \be{eq7}
 3M_{pl}^2H^2=\rho_m+\rho_\Lambda,
 \ee
 where $\rho_m$ and $\rho_\Lambda$ are the energy density of the
 pressureless matter and the holographic dark energy respectively.
 Thus, it is easy to find that
 \be{eq8}
 E(z)=\left[\frac{\Omega_{m0}(1+z)^3}{1-
 \Omega_\Lambda}\right]^{1/2},
 \ee
 where $\Omega_i\equiv\rho_i/(3M_{pl}^2H^2)$ is the fractional
 energy density; the subscript ``0'' indicates the present value
 of the corresponding quantity; $a=(1+z)^{-1}$ (we set $a_0=1$).
 Following~\cite{r29,r43}, from Eq.~(\ref{eq1}) and replacing
 $L$ with the future event horizon $R_h$ in Eq.~(\ref{eq3}),
 we have
 \be{eq9}
 \int_a^\infty\frac{d\ln\tilde{a}}{H\tilde{a}}
 =\frac{c}{Ha\sqrt{\Omega_\Lambda}}\,.
 \ee
 From Eq.~(\ref{eq8}), we obtain
 \be{eq10}
 \frac{1}{Ha}=
 \sqrt{a(1-\Omega_\Lambda)}\frac{1}{H_0\sqrt{\Omega_{m0}}}\,.
 \ee
 Substituting it into Eq.~(\ref{eq9}), we find that
 \be{eq11}
 \int_x^\infty e^{\tilde{x}/2}\sqrt{1-\Omega_\Lambda}d\tilde{x}
 =c\,e^{x/2}\sqrt{\frac{1}{\Omega_\Lambda}-1},
 \ee
 where $x\equiv\ln a$. Taking derivative with respect to $x$ in
 both side of Eq.~(\ref{eq11}) and noting that $a=(1+z)^{-1}$,
 we finally obtain that~\cite{r29,r43}
 \be{eq12}
 \frac{d\Omega_\Lambda}{dz}=-(1+z)^{-1}\Omega_\Lambda
 (1-\Omega_\Lambda)\left(1+\frac{2}{c}\sqrt{\Omega_\Lambda}\right).
 \ee
 One can get $\Omega_\Lambda(z)$ by solving this differential
 equation with the initial condition
 $\Omega_{\Lambda 0}=1-\Omega_{m0}$. Substituting the
 $\Omega_\Lambda(z)$ into Eqs.~(\ref{eq8}) and~(\ref{eq5}),
 the dimensionless age parameter of our universe $T_z(z)$
 is in hand. Then, we can compare $T_z(z)$ with the
 $T_{obj}$ of the three OHRO. It is worth noting that from
 Eqs.~(\ref{eq12}), (\ref{eq8}) and~(\ref{eq5}), $T_z(z)$
 is independent of the Hubble constant $H_0$. On the other
 hand, $T_{obj}$ is proportional to the Hubble constant $H_0$.
 The lower $H_0$, the smaller $T_{obj}$ is.

 \begin{table}[htbp]
 \begin{center}
 \begin{tabular}{c|c|c|c} \hline\hline
 $\ (c,\Omega_{m0})\ $ & $\ \tau (3.91)\ $
 & $\ \tau (1.43)\ $ & $\ \tau (1.55)\ $ \\ \hline
 $(0.73,0.26)$ & 0.918 & 1.276 & 1.362 \\
 $(1.17,0.26)$ & 0.906 & 1.234 & 1.320 \\
 $(0.73,0.32)$ & 0.829 & 1.158 & 1.236 \\
 $(1.17,0.32)$ & 0.820 & 1.128 & 1.205 \\
 \hline\hline
 \end{tabular}
 \end{center}
 \caption{\label{tab1} The ratio $\tau (z)\equiv T_z(z)/T_{obj}$
 at $z=3.91$, $1.43$ and $1.55$, for different model parameters
 $c$ and $\Omega_{m0}$, in the case of SNIa+CMB+LSS without
 prior on $h$.}
 \end{table}

In~\cite{r43}, the original holographic dark energy has been
 constrained recently by combining the latest SNe~Ia, galaxy
 clustering and CMB anisotropy. In the case of SNIa+CMB+LSS
 without prior on $h$, the fit values are $c=0.91_{-0.18}^{+0.26}$
 and $\Omega_{m0}=0.29_{-0.03}^{+0.03}$, while the best fit value
 of $h$ is $0.63$. In the case of SNIa+CMB+LSS with prior
 $h=0.72\pm 0.08$ (which is the final result of
 Freedman {\it et al.}~\cite{r64}), the fit values are
 $c=0.91_{-0.19}^{+0.23}$ and $\Omega_{m0}=0.29\pm 0.03$.
 In the case of SNIa+CMB+LSS with prior
 $0.64\leq h\leq 0.80$, the fit values are
 $c=0.82_{-0.13}^{+0.11}$ and $\Omega_{m0}=0.28_{-0.02}^{+0.03}$.
 We will examine the age problem in these three cases one by one.
 In the case of SNIa+CMB+LSS without prior on $h$, the
 dimensionless age parameter of the OHRO at $z=3.91$, $1.43$
 and $1.55$ are $T_{obj}=0.129$, $0.256$ and $0.226$ respectively,
 for the best fit $h=0.63$. In Table~\ref{tab1}, we present the
 ratio $\tau (z)\equiv T_z(z)/T_{obj}$ at $z=3.91$,
 $1.43$ and $1.55$, for the four combinations of model parameters
 $c$ and $\Omega_{m0}$ of the fit values with $1\,\sigma$
 uncertainty. Obviously, $T_z(z)> T_{obj}$ holds at
 $z=1.43$ and $1.55$, whereas $T_z(z)< T_{obj}$ at $z=3.91$.
 The old quasar APM 08279+5255 at $z=3.91$ cannot be accommodated.
 In the other two cases with prior $h=0.72\pm 0.08$ and
 $0.64\leq h\leq 0.80$, we present the similar contents in
 Tables~\ref{tab2} and~\ref{tab3} respectively. As mentioned above,
 $T_{obj}$ is proportional to the Hubble constant $H_0$. For the
 lower bound $h=0.64$, the dimensionless age parameter of the OHRO
 at $z=3.91$, $1.43$ and $1.55$ are $T_{obj}=0.131$, $0.262$ and
 $0.229$ respectively. From Tables~\ref{tab2} and~\ref{tab3},
 even for the lower bound of $T_{obj}$, we find again that the
 old quasar APM 08279+5255 at $z=3.91$ cannot be accommodated
 in these two cases.

 \begin{table}[htbp]
 \begin{center}
 \begin{tabular}{c|c|c|c} \hline\hline
 $\ (c,\Omega_{m0})\ $ & $\ \tau (3.91)\ $
 & $\ \tau (1.43)\ $ & $\ \tau (1.55)\ $ \\ \hline
 $(0.72,0.26)$ & 0.904 & 1.257 & 1.342 \\
 $(1.14,0.26)$ & 0.892 & 1.217 & 1.302 \\
 $(0.72,0.32)$ & 0.817 & 1.141 & 1.217 \\
 $(1.14,0.32)$ & 0.808 & 1.112 & 1.188 \\
 \hline\hline
 \end{tabular}
 \end{center}
 \caption{\label{tab2} The same as in Table~\ref{tab1}, but
 for the case of SNIa+CMB+LSS with prior $h=0.72\pm 0.08$.}
 \end{table}

 \begin{table}[htbp]
 \begin{center}
 \begin{tabular}{c|c|c|c} \hline\hline
 $\ (c,\Omega_{m0})\ $ & $\ \tau (3.91)\ $
 & $\ \tau (1.43)\ $ & $\ \tau (1.55)\ $ \\ \hline
 $(0.69,0.26)$ & 0.905 & 1.260 & 1.345 \\
 $(0.93,0.26)$ & 0.898 & 1.236 & 1.321 \\
 $(0.69,0.31)$ & 0.830 & 1.161 & 1.238 \\
 $(0.93,0.31)$ & 0.825 & 1.142 & 1.219 \\
 \hline\hline
 \end{tabular}
 \end{center}
 \caption{\label{tab3} The same as in Table~\ref{tab1}, but
 for the case of SNIa+CMB+LSS with prior $0.64\leq h\leq 0.80$.}
 \end{table}

\begin{center}
 \begin{figure}[htbp]
 \centering
 \includegraphics[width=0.5\textwidth]{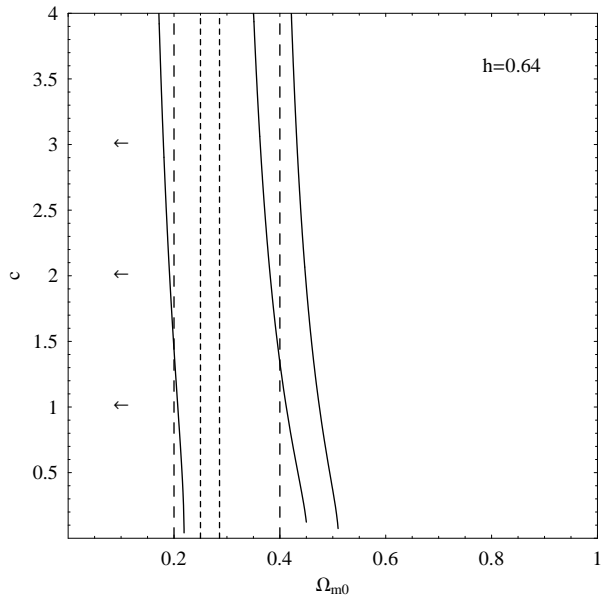}
 \caption{\label{fig1} The three solid lines are,
 from left to right, contours
 $T_z(3.91)=T_{obj}(3.91)$, $T_z(1.43)=T_{obj}(1.43)$ and
 $T_z(1.55)=T_{obj}(1.55)$. The WMAP3 bound
 $\Omega_{m0}=0.268\pm 0.018$~\cite{r6} is indicated by two
 short-dashed lines. The model-independent cluster estimate
 $\Omega_{m0}=0.3\pm 0.1$~\cite{r63} is indicated by two
 long-dashed lines. The allowed parameter spaces are the
 left regions of these contours, as indicated by the arrows.
 In which, we have used the reduced Hubble constant $h=0.64$.}
 \end{figure}
 \end{center}

Let us examine this age problem from another perspective.
 Following~\cite{r50}, we keep the model parameters $c$ and
 $\Omega_{m0}$ free, and find out the parameter space which can
 accommodate these three OHRO by means of plotting the contours
 of the dimensionless age parameter. We scan the parameters in
 the ranges $0<\Omega_{m0}\leq 1$ and $0<c\leq 4$. Note that $c$
 and $\Omega_{m0}$ cannot be zero in order to avoid divergence
 in Eqs.~(\ref{eq12}) and~(\ref{eq5}) with Eq.~(\ref{eq8}).
 And then, we obtain three contours $T_z(3.91)=T_{obj}(3.91)$,
 $T_z(1.43)=T_{obj}(1.43)$ and $T_z(1.55)=T_{obj}(1.55)$.
 As mentioned above, $T_{obj}$ is proportional to the Hubble
 constant $H_0$ whereas $T_z(z)$ is independent of $H_0$.
 Thus, we take $h=0.64$, which is the lower bound of the final
 result $h=0.72\pm 0.08$ of Freedman {\it et al.}~\cite{r64}.
 We present these contours in Fig.~\ref{fig1}. The allowed
 parameter spaces are the left regions of these contours,
 as indicated by the arrows. The WMAP3 bound
 $\Omega_{m0}=0.268\pm 0.018$~\cite{r6} is also indicated by
 two short-dashed lines. It is easy to see that the OHRO at
 $z=1.43$ and $1.55$ can be accommodated. However, the OHRO
 at $z=3.91$ cannot be accommodated, since the allowed
 parameter space are out of the WMAP3 bound. Even when we use
 instead the loosest and model-independent cluster estimate
 $\Omega_{m0}=0.3\pm 0.1$~\cite{r63} which is indicated by
 two long-dashed lines, the allowed parameter space becomes
 very narrow for the OHRO at $z=3.91$. This narrowed
 parameter space $0.2\leq\Omega_{m0}\,\lsim\, 0.22$, however,
 can also be ruled out by the combined constraints from the
 latest SNe~Ia, galaxy clustering and CMB anisotropy~\cite{r43}.
 Therefore, the old quasar APM 08279+5255 at $z=3.91$ cannot
 be accommodated in the original holographic dark energy model.
 The age problem also exists, like in the other dark energy
 models mentioned in Sec.~\ref{sec1}.


\section{Alleviating the age problem by lower Hubble constant}\label{sec3}
To solve the age problem, one way is to decrease the Hubble
 constant. This can be seen from Eq.~(\ref{eq6}). As mentioned
 above, $T_z(z)$ is independent of the Hubble constant $H_0$,
 whereas $T_{obj}$ is proportional to the Hubble constant $H_0$.
 The lower $H_0$, the smaller $T_{obj}$ is. Therefore, the
 condition Eq.~(\ref{eq6}) can be satisfied more easily. In the
 previous analysis, the reduced Hubble constant $h=0.72\pm 0.08$
 of Freedman {\it et al.}~\cite{r64} has been used extensively.
 However, in the recent years, it is argued that there is
 systematic bias in the result of Freedman {\it et al.}~\cite{r64}.
 Sandage and collaborators advocate a lower Hubble constant in
 a series of works~\cite{r65,r66,r67,r68,r69}. Their final result
 reads $h=0.623\pm 0.063$~\cite{r69}.

\begin{center}
 \begin{figure}[htbp]
 \centering
 \includegraphics[width=0.5\textwidth]{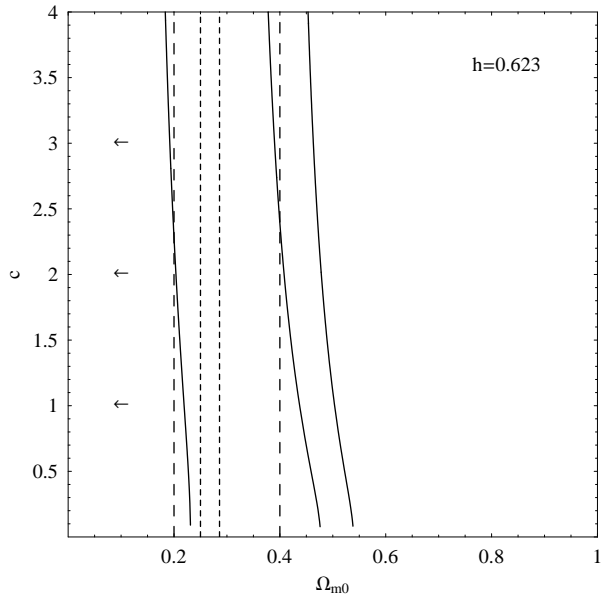}
 \caption{\label{fig2} The same as in Fig.~\ref{fig1},
 except for $h=0.623$.}
 \end{figure}
 \end{center}

At first, we take the central value of the result of
 Sandage {\it et al.}~\cite{r69}, namely $h=0.623$. We present
 the contours $T_z(3.91)=T_{obj}(3.91)$, $T_z(1.43)=T_{obj}(1.43)$
 and $T_z(1.55)=T_{obj}(1.55)$ in Fig.~\ref{fig2}. Obviously,
 the situation is similar to the case $h=0.64$ considered in the
 previous section, although there are some slight improvements.
 The OHRO at $z=1.43$ and $1.55$ are easily accommodated. However,
 the old quasar APM 08279+5255 at $z=3.91$ still cannot be
 accommodated in the original holographic dark energy model. This
 is mainly because the difference between $h=0.623$ and $h=0.64$
 is too small.

\begin{center}
 \begin{figure}[htbp]
 \centering
 \includegraphics[width=0.5\textwidth]{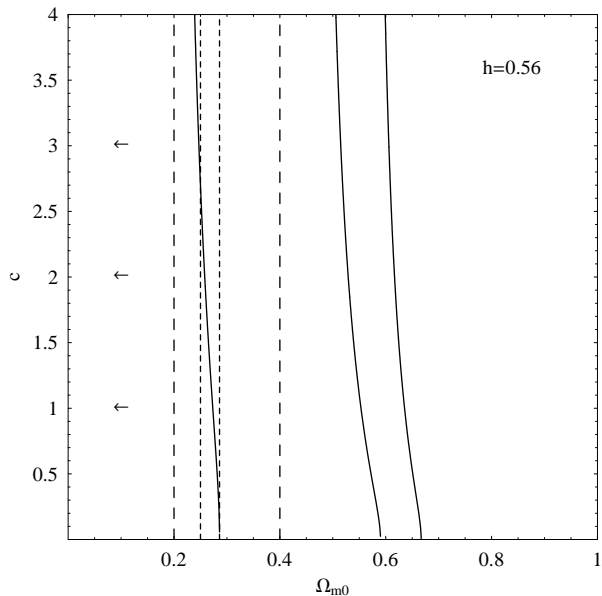}
 \caption{\label{fig3} The same as in Fig.~\ref{fig1},
 except for $h=0.56$.}
 \end{figure}
 \end{center}

Then, we consider the lower bound of the result of
 Sandage {\it et al.}~\cite{r69}, i.e. $h=0.56$. The corresponding
 contours are presented in Fig.~\ref{fig3}. It is easy to see
 that the situation is improved significantly. The allowed
 parameter space for the OHRO at $z=3.91$ is fully consistent with
 the loosest model-independent cluster estimate
 $\Omega_{m0}=0.3\pm 0.1$~\cite{r63}, the tighter WMAP3 bound
 $\Omega_{m0}=0.268\pm 0.018$~\cite{r6}, and the combined
 constraints from the latest SNe~Ia, galaxy clustering and CMB
 anisotropy~\cite{r43}. In this case, all the three OHRO at
 $z=3.91$, $1.43$ and $1.55$ can be accommodated in the original
 holographic dark energy model. The age problem disappears.


\section{Concluding remarks}\label{sec4}
Actually, in addition to the three OHRO considered in this work,
 there are other OHRO in the literature, for instance, the 4.0~Gyr
 old radio galaxy 3C~65 at $z=1.175$~\cite{r70}, and the high
 redshift quasar B1422+231 at $z=3.62$ whose best fit age is
 1.5~Gyr with a lower bound of 1.3~Gyr~\cite{r71}. However, they
 cannot be used to constrain the model parameters as restrictive
 as the previous three OHRO. Thus, we do not consider them in
 this work.

In this note, we test the original holographic dark energy model
 with some old high redshift objects. The main idea is very
 simple: the universe cannot be younger than its constituents.
 We find that the original holographic dark energy model can be
 ruled out, unless a lower Hubble constant is taken.

In fact, as mentioned in Sec.~\ref{sec1}, the old quasar
 APM 08279+5255 at $z=3.91$ cannot be accommodated in many
 dark energy models (including the $\Lambda$CDM model). The
 common way out is to use a lower Hubble constant instead.
 This hints that the extensively used $h=0.72\pm 0.08$ of
 Freedman {\it et al.}~\cite{r64} may have systematic bias.
 The ages of OHRO tend to a lower Hubble constant, say, the
 one recently advocated by Sandage and collaborators.

In the literature, there are still some debates on the value
 of the Hubble constant. After the final result $h=0.72\pm 0.08$
 of Freedman {\it et al.}~\cite{r64}, many authors argue for
 a lower Hubble constant, for instance, $h=0.68\pm 0.07$
 at $2\,\sigma$ uncertainty in~\cite{r72}. By now, the final
 result $h=0.623\pm 0.063$ of Sandage and
 collaborators~\cite{r69} attracted more and more attentions.
 We consider that it is important to take this into account
 when one tries to constrain the cosmological model parameters.
 The age problem in dark energy models supports this argument
 strongly.

It is worth noting that the results in this work are obtained
 in the original holographic dark energy model. In fact, there
 are many modified holographic dark energy models in the
 literature, such as the nonsaturated holographic dark
 energy~\cite{r73}, the holographic dark energy model with
 variable $G_N$~\cite{r74}, the interacting holographic dark
 energy model~\cite{r75,r76}, and so on. The conclusions of
 this work might be changed in these models. For instance,
 in the interacting holographic dark energy model, it is shown
 that for a fixed $c$, the holographic dark energy starts
 to be effective earlier and consequently the universe
 experiences the accelerated expansion earlier when the
 interaction is larger~\cite{r76}. In this case, the universe
 can have a longer age. Therefore, it is of interest to
 examine the age problem in these modified holographic
 dark energy models in the future works.


\section*{ACKNOWLEDGMENTS}
We are grateful to Prof.~Rong-Gen~Cai, Prof.~Miao~Li and
 Prof.~Bin~Wang for helpful discussions. We also thank
 Shi-Chao~Tang, Jian~Hu, Xin~Liu, Yue~Shen, Lin~Lin, Sumin~Tang,
 Jing~Jin, Wei-Ke~Xiao, Feng-Yun~Rao, Nan~Liang, Rong-Jia~Yang,
 Jian~Wang, Yuan~Liu, and Xin~Zhang, Zong-Kuan~Guo, Qing-Guo~Huang,
 Hui~Li, Meng~Su for kind help and discussions. The major computations
 of this work were completed during the period of the International
 Conference on Astrophysics of Compact Objects, Huangshan, China,
 July 2007. We acknowledge partial funding support from China
 Postdoctoral Science Foundation, and by the Ministry of Education of
 China, Directional Research Project of the Chinese Academy of Sciences
 under project No.~KJCX2-YW-T03, and by the National Natural Science
 Foundation of China under project No.~10521001.

 \renewcommand{\baselinestretch}{1.1}


\end{document}